\documentclass{article}

\usepackage[final, nonatbib]{neurips_2023_ml4ps}

\usepackage[utf8]{inputenc} 
\usepackage[T1]{fontenc}    
\usepackage{hyperref}       
\usepackage{url}            
\usepackage{booktabs}       
\usepackage{tabularx}
\usepackage{amsfonts}       
\usepackage{nicefrac}       
\usepackage{microtype}      
\usepackage{wrapfig}
\usepackage[dvipsnames]{xcolor}

\usepackage{amssymb}
\usepackage{pifont}
\newcommand{\cmark}{\ding{51}}%
\newcommand{\xmark}{\ding{55}}%
\usepackage[backend=biber,natbib=true, sorting=none]{biblatex}
\usepackage{graphicx}

\title{
High-Cadence Thermospheric Density Estimation enabled by Machine Learning on Solar Imagery}

\author{%
  Shreshth A. Malik\thanks{Equal contribution.}\\
  University of Oxford\\
  \And
  James Walsh$^{*}$ \\
  University of Cambridge\\
  \And
  Giacomo Acciarini \\
  University of Surrey\\
  \And
  Thomas E. Berger\\
  University of Colorado at Boulder
    \And
    Atılım Güneş Baydin\\
    University of Oxford
}

\begin{document}

\maketitle

\begin{abstract}
Accurate estimation of thermospheric density is critical for precise modeling of satellite drag forces in low Earth orbit (LEO). Improving this estimation is crucial to tasks such as state estimation, collision avoidance, and re-entry calculations. The largest source of uncertainty in determining thermospheric density is modeling the effects of space weather driven by solar and geomagnetic activity. Current operational models rely on ground-based proxy indices which imperfectly correlate with the complexity of solar outputs and geomagnetic responses. In this work, we directly incorporate NASA's Solar Dynamics Observatory (SDO) extreme ultraviolet (EUV) spectral images into a neural thermospheric density model to determine whether the predictive performance of the model is increased by using space-based EUV imagery data instead of, or in addition to, the ground-based proxy indices. We demonstrate that EUV imagery can enable predictions with much higher temporal resolution and replace ground-based proxies while significantly increasing performance relative to current operational models. 
Our method paves the way for assimilating EUV image data into operational thermospheric density forecasting models for use in LEO satellite navigation processes.
\end{abstract}


\section{Introduction}
\label{sec:intro}
As the number of space objects in the low-Earth orbit (LEO) grows \cite{mcdowell2020low}, there is a need for a more accurate model of the thermospheric density variability in order reduce uncertainty in trajectory calculations and improve the accuracy of collision probabilities. The largest source of uncertainty in LEO satellite trajectory calculations is the drag force, which in turn is driven largely by uncertainties in thermospheric density estimation \cite{berger2020flying}. Currently, the large uncertainties characteristic of operational thermospheric density models translate directly into more frequent conjunction warnings as resident space object trajectory covariances frequently exceed warning thresholds. A major goal of modern space traffic coordination (STC) is to reduce the number of low-probability conjunction warnings issued to satellite operators and thus reduce their workload in adjudicating and mitigating possible collisions. Hence, it is crucial to reduce the uncertainties in thermospheric density models as well as to improve the tracking frequency and characterization of satellites and debris in LEO. 

Basal thermospheric density levels are primarily set by heating from the absorption of extreme ultraviolet (EUV) photons from the Sun. In addition, geomagnetic activity due to the interaction of solar wind transients and solar Coronal Mass Ejections (CMEs) with the Earth's magnetosphere can cause impulsive heating and large deviations from basal levels, particularly during geomagnetic storms. Current operational empirical models rely on simple daily proxy indices to represent the basal and transient solar impacts to the thermosphere, e.g. the F10.7 proxy index \cite{tapping201310} which imperfectly correlates with true EUV output, and the Kp or Ap indices \cite{menvielle2010geomagnetic} to represent geomagnetic activity. Both indices are fairly crude representations of physical interactions, lacking in both spatial and temporal resolution that results in underestimating the dynamics of the solar-terrestrial system. The High Accuracy Satellite Drag Model (HASDM) is the operational thermospheric density model used by the US Space Force (USSF) for their STC mission and is based on the empirical JB08 density model \cite{bowman2008new}. HASDM compensates the shortcomings due to use of proxy indices in the JB08 model by deriving temperature corrections driven by assimilation of calibration satellite trajectory data. However the model does not forecast well during high solar activity periods due to the limitations of using proxy data inputs. For some time now, real-time EUV irradiance data has been available from NOAA weather satellites and from NASA research satellites. It is therefore interesting to investigate whether the use of this high-cadence data in a predictive model would improve density estimation during high solar activity, e.g., during solar flare events that can increase the EUV and X-ray irradiance of the Sun by an order of magnitude over time periods as short as an hour.

In this work, we directly incorporate data from NASA's Solar Dynamics Observatory (SDO) EUV images \cite{lemen2012atmospheric, scherrer2012helioseismic} into a neural thermospheric density model to investigate and quantify how additional information of the EUV spectrum and the solar activity can benefit the prediction of high-cadence thermospheric density variations \cite{vourlidas2018euv}. 
We demonstrate that compressed latent space embeddings of SDO imagery generated by a variational autoencoder can replace ground-based proxies while improving predictive accuracy and temporal resolution. This provides a pathway to near real-time, high fidelity density estimation.

\section{Background and Related Work}
\label{sec:bg}

\paragraph{Drivers of thermospheric density change}
The dynamic changes in the thermospheric density are caused by various space weather processes including EUV irradiance from the Sun and energy deposition from the Earth's magnetosphere during geomagnetic storm periods \cite{knipp2004direct}.  EUV photons are deposited in the Earth's upper atmosphere at thermospheric heights (200--1000 km above sea level) where they provide the majority of the baseline energy input and heating. During geomagnetic storms, thermospheric density can rapidly change by tens of percent on top of the baseline set by the solar irradiance with concomitant changes in the drag force experienced by LEO resident space objects. 
The position of the Earth relative to the Sun, as well as the Earth's rotation cause diurnal and seasonal variations of the EUV irradiance, and short term solar variation from solar flares (the electromagnetic radiation following magnetic eruptions in the solar outer atmosphere) can last 1--4 hours and cause density perturbations of up to 20\% \cite{le2015global}.

\paragraph{Existing thermospheric density models} Full physics-based models, while highly accurate, are computationally intractable for real-time operations \cite{qian2014ncar}. Instead, operational (empirical)
models such as JB08 \cite{bowman2008new}, MSIS \cite{picone2002nrlmsise} and HASDM \cite{storz2005high} are simplified physics-constrained functions that have been fit to observational data. However, these rely on proxy indices to represent solar EUV irradiance. While computationally efficient, proxies imperfectly capture complex spatiotemporal solar dynamics, limiting density prediction accuracy and cadence to daily predictions. 
Recent machine-learning-based approaches  \cite{licata2022uncertainty, turner2020machine} such as Karman\footnote{\url{https://github.com/spaceml-org/karman}, date of access: September 2023.} \cite{bonasera2021dropout, acciarini-2023-karman} demonstrate improved performance over operational models. However, while there has been an addition of FISM2 EUV irradiance Stan bands in Karman, these models still do not account for the full complexity of the solar activity. 

\section{Data}

Here we outline the input features and density targets used in this work. For more information, a detailed table is provided in the Appendix \ref{sec:app-data}.

\paragraph{SDO data, solar proxies and geomagnetic indices}

The Solar Dynamics Observatory provides detailed solar imagery from the Atmospheric Imaging Assembly (AIA) \cite{lemen2012atmospheric}, EUV Variability Experiment (EVE), and Helioseismic and Magnetic Imager (HMI) \cite{scherrer2012helioseismic} instruments. The AIA and HMI data are incorporated through a machine-learning-ready derivative dataset, SDOML \cite{galvez2019machine}, containing 512x512 imagery spanning two ultraviolet and seven extreme ultraviolet wavelengths, as well as magnetic vector field components $B_x$, $B_y$, $B_z$, resulting in a 12 channel image input (example in Appendix \ref{sec:app-data}). Data is currently available at a 12 minute time resolution.  
Empirical models use solar proxy indices: F10.7, M10.7, S10.7, \& Y10.7 \cite{vourlidas2018euv}, and geomagnetic indices: Kp, Ap \& Dst \cite{menvielle2010geomagnetic} to account for geomagnetic storm activity, which are all provided daily.

\paragraph{Thermospheric density ground truth}
The primary source of ground truth for training tasks is TUDelft’s thermosphere data \cite{doornbos2012thermospheric}. These data contains precise orbit determination-derived thermospheric drag values for CHAMP, GRACE, GRACE-FO, GOCE, and Swarm missions \cite{doornbos2012thermospheric, siemes2016swarm, van2020thermosphere}. It contains satellite time, altitude, and location information as well which is used by models for prediction. The data was sourced from multiple providers, with observations occurring at different time intervals. GOCE \& Swarm at 10 seconds, CHAMP \& GRACE recorded once per 30 seconds. We removed orbits with outlier density measurements caused by missing data (Appendix \ref{sec:app-data}).

\paragraph{Dataset preprocessing}
The dataset was aligned based on a forward-nearest fill policy. All experiments were carried out at a 12-minute cadence, beginning on 4 May 2010, and iterated forward until 29 November 2018. Following Karman \cite{acciarini-2023-karman}, non-SDO inputs to models were 0/1 normalized and cyclical features were produced for spatiotemporal inputs. The validation and test subsets were selected by choosing one month of the calendar year, changing per year to the following month. 

\section{Methods} 
\label{sec:methods}

We follow a three step method for investigating whether EUV imagery can be used for accurate thermospheric density modeling. 
First we generate general-purpose compressed embeddings of the imagery. Then we add these features as input into a physics-informed deep learning model. Finally, we systematically evaluate the performance of models with and without solar proxies and SDO data in an ablation study. For more details on models, please refer to Appendix \ref{sec:app-models}.

\paragraph{Solar embeddings}
We first extract valuable features from SDOML data. Building on prior work \cite{brown2021}, we employ a variational autoencoder (VAE) trained to reconstruct SDOML data from a lower-dimensional encoding, offering a highly compressed representation of the Sun's state for downstream tasks, such as thermospheric density prediction. 
Our VAE architecture is also enhanced with residual blocks \cite{he2016deep}, enabling variable latent dimension testing \cite{zhu2017unpaired}. 

\begin{figure}[t]
  \centering
  \includegraphics[width=\textwidth]{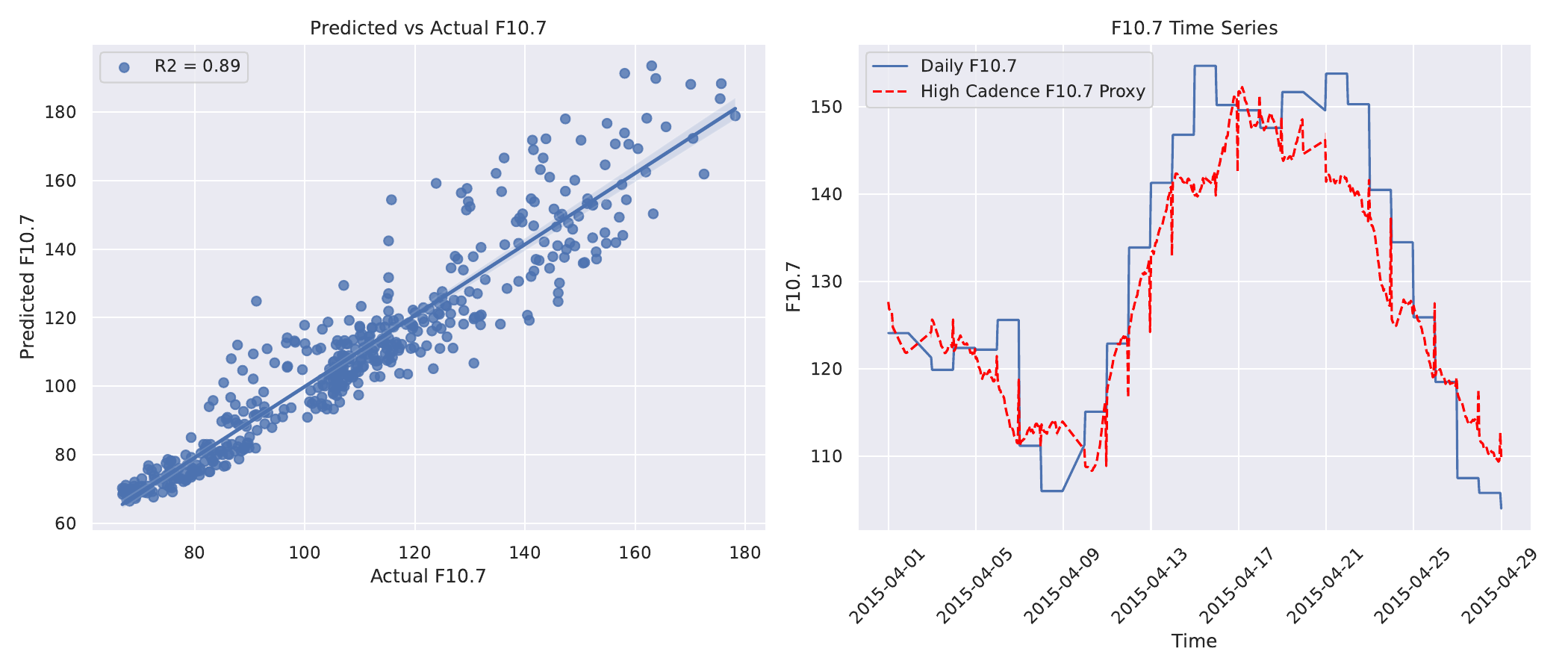}
  \caption{F10.7 predictions on test set against ground truth. Parity plot between averaged daily F10.7 predictions from the model against the ground truth (left). Time series of F10.7 values and corresponding high-cadence model predictions (right).}
  \label{fig:f10}
\end{figure}

\begin{wrapfigure}{r}{0.5\textwidth}
  \begin{center}
    \includegraphics[width=0.48\textwidth]{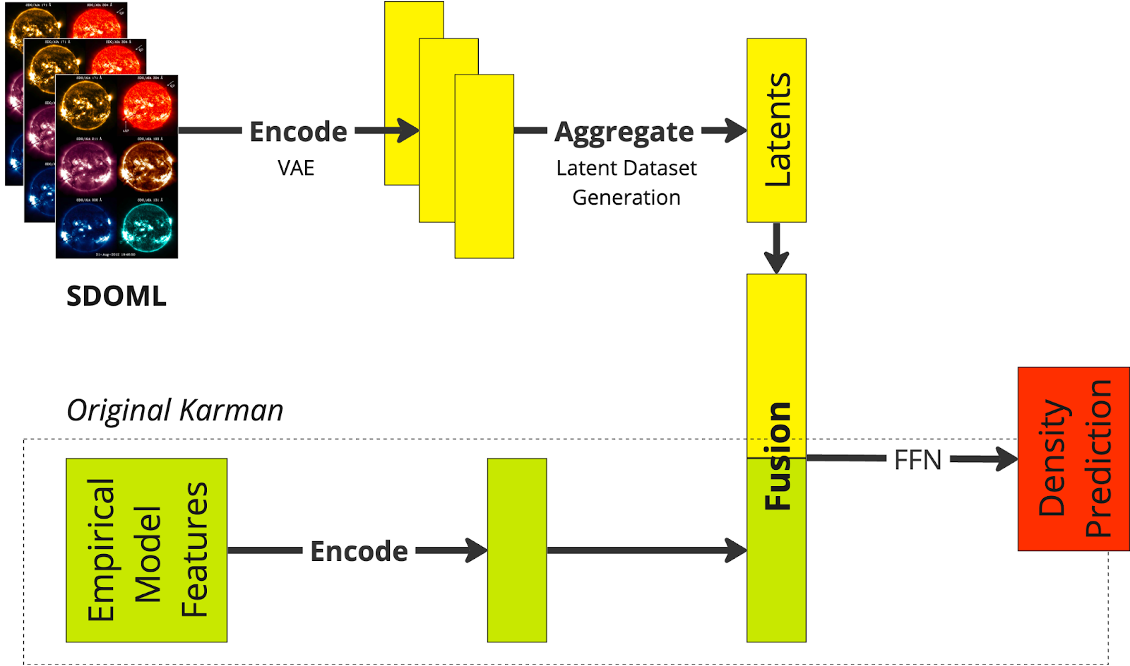}
  \end{center}
  \caption{Model pipeline diagram. SDOML images are encoded using the VAE (Section \ref{sec:methods}), aggregated, then concatenated with the other input features in a feed forward neural network (FFN).}
  \label{fig:model}
  \vspace{-1em}
\end{wrapfigure}

\paragraph{SDO thermospheric density model}
In Karman \cite{acciarini-2023-karman}, a simple feed-forward network (FFN) was used to process the input empirical model features. Here we additionally embed SDO imagery with the VAE as described and mean-aggregate embeddings from a 4 hour history at a 12 minute resolution for prediction. We concatenate this SDO time series embedding with the empirical model feature embedding and process this through a further FFN for density prediction (Figure \ref{fig:model}).

\paragraph{Physics-informed residual learning}
Karman \cite{acciarini-2023-karman} directly predicted the (log) density from input features. Here, we also investigated predicting residuals to a physics-informed model. We use an exponential atmosphere (EXP) here \cite{vallado2001fundamentals}\footnote{\url{https://github.com/lcx366/ATMOS/blob/master/pyatmos/standardatmos/expo.py} date of access: September 2023}, which captures the simplest physics, but in principle any baseline can be used, such as another empirical model. This allows the model to focus on correcting the model for the additional information provided by the SDOML images while staying close to the underlying physics.
Density predictions vary by orders of magnitude with altitude, so we use residuals in the form: $\textrm{R}(x, y)=\textrm{log}\left(y\right) - \textrm{log}\left(\textrm{EXP}(x)\right)$, where $x$ are the input features and $y$ is the true density, as targets to train our network. 
We plan to include more physical inductive biases in the future, such as smoothness in space and time.

\begin{figure}
  \centering
  \includegraphics[width=\textwidth]{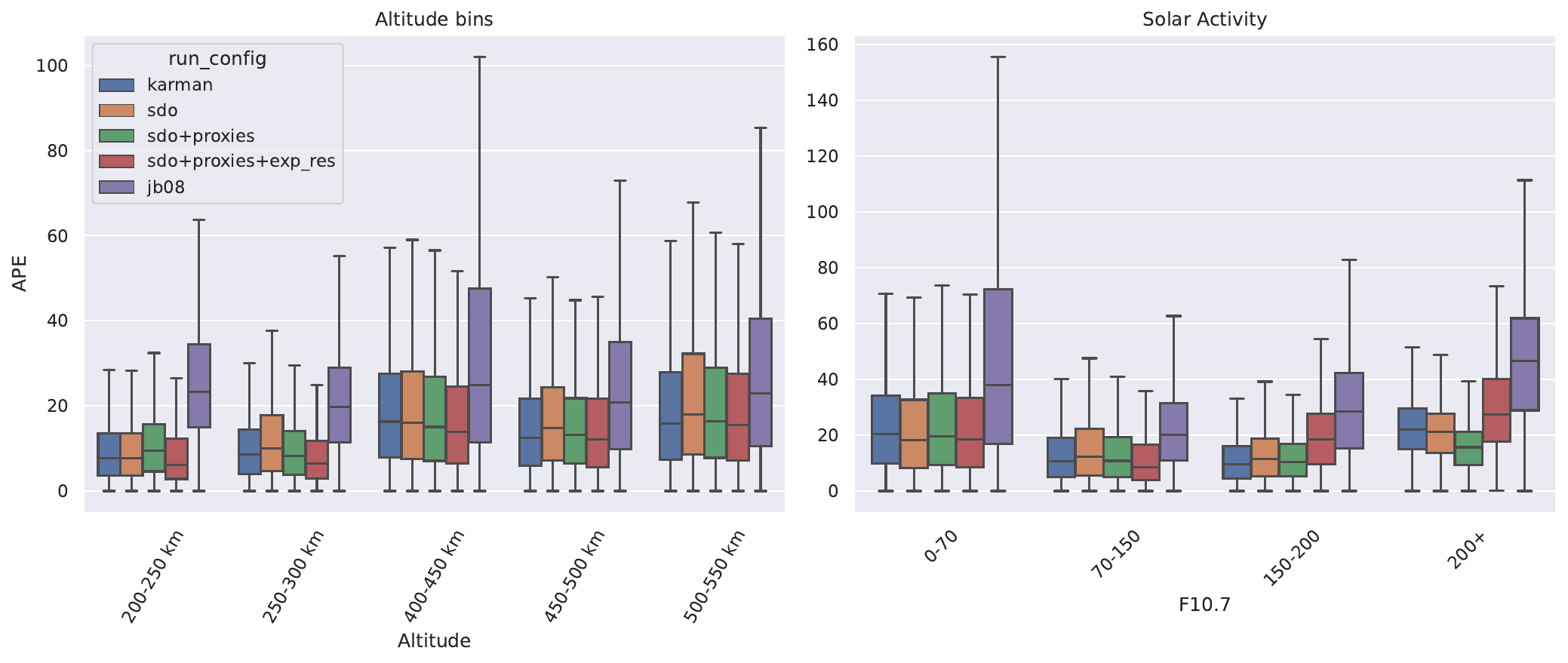}
  \caption{Ablation study model performance. Distribution of absolute percentage error (APE) for each model across altitude ranges (left), and solar activity levels (right).}
  \label{fig:results}
\end{figure}

\section{Results}
\label{sec:results}

We show that the inclusion of direct EUV imagery not only allows for higher-cadence reconstruction of solar indices to use with existing models, but also accurate end-to-end density prediction.

\paragraph{Solar embeddings can reconstruct solar indices at high temporal resolution.} 
To validate the embeddings provide relevant information on solar activity, we use the VAE latent means as inputs into a simple feed-forward neural network (Appendix \ref{sec:app-models}) to predict the F10.7 solar index. We find that we are able to predict F10.7 at a high accuracy using just SDO embeddings, with a root mean squared error (RMSE) of \textbf{8.32}. This is significantly better than random embeddings (RMSE = 32.7).
Note this is not an auto-regressive model. 
If we were to include previous F10.7 values, we would expect performance to increase even further. 
A key byproduct of this investigation is that we now have a higher cadence ML-informed replacement for F10.7 for use in existing operational models. F10.7 is only given daily but SDOML imagery is available every 12 minutes, so we can also predict inter-day values (Figure \ref{fig:f10}). 

\begin{table}[]
  \caption{Ablation study results on test fold. The mean absolute percentage error (MAPE), mean absolute error (MAE), root mean squared error (RMSE), and median symmetric accuracy (MSA \cite{morley2018measures}) is shown. Lower is better for all. }
  \label{tab:results}
\begin{tabular}{llllllll}
\toprule
\textbf{Model} &
   \textbf{\begin{tabular}[t]{@{}l@{}}Solar \\ Imagery\end{tabular}}&
  \textbf{\begin{tabular}[t]{@{}l@{}}Solar \\ Proxies\end{tabular}} &
  \textbf{\begin{tabular}[t]{@{}l@{}}Physics\\ informed\end{tabular}} &
  \textbf{MAPE} &
  \textbf{\begin{tabular}[t]{@{}l@{}}MAE\\/ $10^{-12}$\end{tabular}} &
  \textbf{\begin{tabular}[t]{@{}l@{}}RMSE\\/ $10^{-12}$\end{tabular}} &
  \textbf{MSA} \\ \midrule \midrule
JB08 \cite{bowman2008new}                           & \color{red}{\xmark} & \color{Green}{\cmark} & \color{Green}{\cmark} & 21.61 & 3.99 & 7.22          & 44.8 \\
MSIS \cite{picone2002nrlmsise}                       & \color{red}{\xmark} & \color{Green}{\cmark} & \color{Green}{\cmark} & 21.54 & 3.04 & 5.82          & 44.7 \\ \hline
Karman \cite{acciarini-2023-karman}                      & \color{red}{\xmark} & \color{Green}{\cmark} & \color{red}{\xmark} & 15.61 & 1.79 & 3.60          & 41.0 \\
Karman - Proxies                 &\color{red}{\xmark}& \color{red}{\xmark} & \color{red}{\xmark} & 23.54 & 2.69 & 5.27          & 43.4 \\
\begin{tabular}[t]{@{}l@{}}Karman - Proxies \\ \: + EXP. Residual\end{tabular} & \color{red}{\xmark} & \color{red}{\xmark} & \color{Green}{\cmark} & 23.45 & 2.30 & 4.85          & 42.8 \\
\textbf{Karman + EXP. Residual} &
  \color{red}{\xmark} &
  \color{Green}{\cmark} &
  \color{Green}{\cmark} &
  14.18 &
  \textbf{1.72} &
  4.32 &
  \textbf{40.2} \\ \midrule
SDO                              & \color{Green}{\cmark} & \color{red}{\xmark} & \color{red}{\xmark} & 16.67 & 1.93 & 3.63          & 41.6 \\
SDO + Proxies                    & \color{Green}{\cmark} & \color{Green}{\cmark} & \color{red}{\xmark} & 15.69 & 1.80 & \textbf{3.56} & 41.1 \\
SDO + EXP. residual              & \color{Green}{\cmark} & \color{red}{\xmark} & \color{Green}{\cmark} & 14.50 & 1.90 & 4.36          & 40.8 \\
\textbf{\begin{tabular}[t]{@{}l@{}}SDO + Proxies \\ \: + EXP. residual\end{tabular}} &
  \color{Green}{\cmark} &
  \color{Green}{\cmark} &
  \color{Green}{\cmark} &
  \textbf{14.00} &
  \textbf{1.72} &
  4.23 &
  \textbf{40.3} \\ \bottomrule
\end{tabular}
\end{table}

\paragraph{Solar embeddings can replace proxies in thermospheric density models.}
Having now validated the efficacy of the embeddings, we utilize them for thermospheric density prediction. We show results for the best performing models on a validation fold across each model type in Table \ref{tab:results}. We find that an ML model using SDO has comparative performance to one using solar proxies (Karman \cite{acciarini-2023-karman}), and greatly outperforms the standard empirical models. Including the proxies as well as SDO data also improves performance. The physics-informed method of predicting residuals to a simple physics model results in the best performance, significantly reducing the error in the baseline Karman model as well. Figure \ref{fig:results} shows error rates for each model across altitudes and solar activities. We find that SDO models work better in periods of high solar activity.

\section{Conclusions}
\label{sec:disc}

\paragraph{Contributions}
In this work, we have shown that we can replace daily ground-based proxies for EUV irradiance such as F10.7 with solar imagery from the SDO mission for thermospheric density predictions. By doing so, we enable accurate, higher cadence, near real-time density estimation for LEO operations. From a science perspective, this shows that SDO imagery contains the relevant information for accurate thermospheric density prediction 
for the first time. 

\paragraph{Further work} 

To ascertain whether our model can effectively capture short-term solar variability such as flares, we should investigate time periods where flares have had a significant impact on density \cite{briand2021solar}. An initial investigation of the ground truth density data found that flare effects were insignificant compared to noise (Appendix \ref{sec:app-short-term}). Thus we must increase the fidelity of the measurements to get the most benefit from the high temporal resolution EUV data. Research into extracting densities from LEO satellite constellations and engineering data will be a promising area to investigate. 

\section*{Acknowledgements}

This work is the research product of FDL-X (fdlxhelio.org) an initiative of Frontier Development Lab (FDL.ai). FDL is a public/private partnership between NASA, Trillium Technologies and commercial AI partners Google Cloud and NVIDIA Corporation.  FDL-X and its outputs have been designed, managed and delivered by Trillium Technologies Inc (trillium.tech). Trillium is a research and development company with a focus on intelligent systems and collaborative communities for planetary stewardship, space exploration and human health.  
We express our gratitude to Google Cloud for providing extensive computational resources, as well as to NVIDIA Corporation for use of DGX Cloud.
The material is based upon work supported by NASA under award No(s) 80ARC018D0010-80GSFC18F0114. Any opinions, findings, and conclusions or recommendations expressed in this material are those of the author(s) and do not necessarily reflect the views of the National Aeronautics and Space Administration. 
FDL aspires to ensure that the latest tools and techniques in Artificial Intelligence (AI) and Machine Learning (ML) are applied to developing open science for all Humankind.

\printbibliography

\newpage

\appendix

\section{Data}
\label{sec:app-data}

\begin{figure}[h]
    \centering
    \includegraphics[width=1\textwidth]{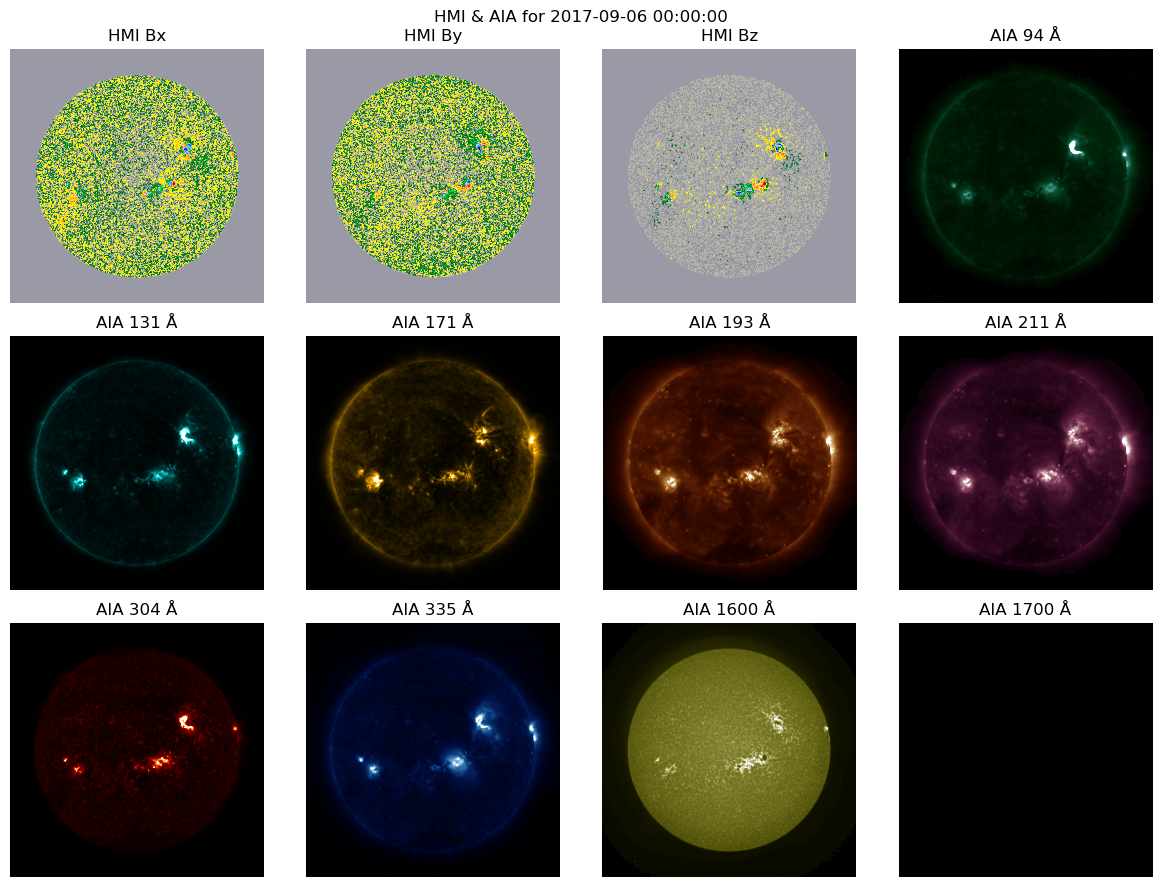}
    \caption{SDO imagery from the SDOML dataset \cite{galvez2019machine}, ordered as three magnetic vector components, seven extreme ultraviolet and two ultraviolet. The final image was unavailable due to sensor protection during geomagnetic storming.}
    \label{fig:sdoml-ground}
\end{figure}

A detailed breakdown of the data sources used in this work is presented in Table \ref{tab:data}. An example of the 12 channel SDO imagery used in this work is shown in Figure \ref{fig:sdoml-ground}.

\paragraph{Data Cleaning}

\begin{figure}
    \centering
    \includegraphics[width=0.5\textwidth]{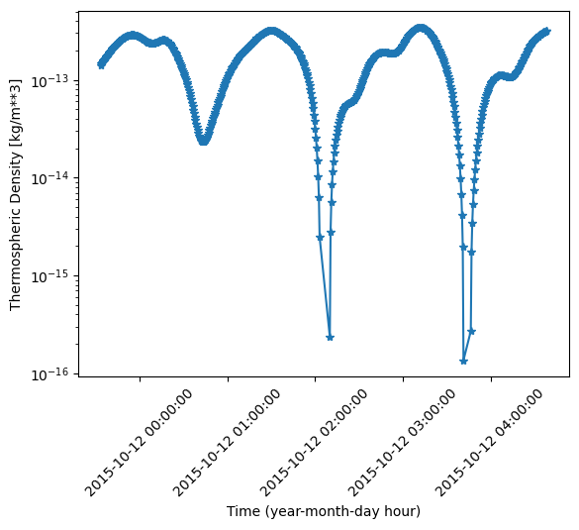}
    \caption{Instance of two orbits whereby a greater than 30 second gap in data occurred, leading to an erroneously low density measurement.}
    \label{fig:data-clean}
\end{figure}

We originally detected the need for a data cleaning process due to high prediction error in limited circumstances. The cleaning process occurred by first detecting periods of time greater than 30 seconds of continuously missing data. These brief instances in 1.05\% of the dataset contributed to examples of sub 10$^{-14}$ kgm$^{-3}$ recordings of density, which are outliers compared to the usual behavior of the thermospheric density across the satellites’ orbits. While we have not processed the precise orbit determination data ourselves, and therefore do not possess all elements to assert if these values were actually the thermospheric density experienced by the satellite during its motion, the lack of data during these periods as well as the fact that these values are significant outliers (i.e., by two orders of magnitude) were enough to invalidate these orbits. An example is provided in Figure \ref{fig:data-clean}, whereby the full orbit would then be rejected from the dataset time series. 

\begin{table}[]
\caption{Overview of data sources used in this work.}
\label{tab:data}
\begin{tabular}{@{}lll@{}}
\toprule
\textbf{Name \& Type}                                                                                                  & \textbf{Description}                                                                                                                                                                                                                                                                                                                                                                                                                                                                                         & \textbf{Granularity \& Source}                                                                                                                                                      \\ \midrule \midrule
 
\begin{tabular}[c]{@{}l@{}}\textbf{SDOML (v2)}\\ Training Data\\ \\ Imagery \\ 512x512/0.6,\\ 512x512/0.5\\ arcsec\end{tabular} & \begin{tabular}[c]{@{}l@{}}Atmospheric Imaging Assembly (AIA; Lemen et al. 2012) \\ 2 ultraviolet, 1600 \& 1700 Å\\ 7 extreme ultraviolet, 94, 131, 171, 193, 211, 304,\\ and 335 Å Helioseismic and Magnetic Imager (HMI; \\ Schou et al. 2012) - visible filtergrams processed into:\\ photospheric Dopplergrams, line-of-sight magnetograms\\ and vector magnetograms\end{tabular}                                                                                                                                                & \begin{tabular}[c]{@{}l@{}}2010-2020\\ 12 second cadence\\ for AIA\\ \\ NASA Solar \\ Dynamics Observatory\\ (SDO) \end{tabular}                                              \\ \midrule
 
\begin{tabular}[c]{@{}l@{}}\textbf{CHAMP}\\ Ground Truth\end{tabular}                                                           & \begin{tabular}[c]{@{}l@{}}CHAMP, a LEO satellite launched on July 15, 2000,\\ orbits Earth in 90 minutes with an initial altitude of\\ 456 km and an orbit inclination of 87.3°. CHAMP\\ needs 131 days to cover all local times\\ (longitudinal cycle). CHAMP re-entered in \\ September 2010. Includes satellite location data.\end{tabular}                                                                                                                                                                                   & \begin{tabular}[c]{@{}l@{}}May 2000-Sep 2010\\ 30 second cadence\\ GFZ\\ \\ TUDelft\end{tabular}                                                                             \\ \midrule
 
\begin{tabular}[c]{@{}l@{}}\textbf{GOCE}\\ Ground Truth\end{tabular}                                                            & \begin{tabular}[c]{@{}l@{}}GOCE is a LEO satellite launched on 17 March 2009\\ at 270km and 96.5° orbit inclination. The mass density\\ estimates were determined from the along-track\\ acceleration of the GOCE satellite’s ion propulsion\\ array with changing thrust. The retrieved GOCE mass\\ densities are made available by the ESA. GOCE\\ re-entered in October 2013. Includes satellite\\ location data.\end{tabular}                                                                                                 & \begin{tabular}[c]{@{}l@{}}Nov 2009-Oct 2013\\ 10 second cadence\\ \\ ESA Earth Online\\ \\ TUDelft\end{tabular}                                                             \\ \midrule
 
\begin{tabular}[c]{@{}l@{}}\textbf{GRACE}/\\ \textbf{GRACE-FO}\\ Ground Truth\end{tabular}                                               & \begin{tabular}[c]{@{}l@{}}GRACE-A and GRACE-B were twin satellites\\ launched in March 2002. GRACE-B kept 220 km\\ from GRACE-A. Both spacecraft derived comparable\\ mass densities. GRACE started at 500 km. GRACE\\ had a 95-minute orbit and 89.5° declination. GRACE\\ longitudinal coverage is about 160 days. GRACE\\ reentered in March 2018. 
\\ GRACE-FO mission maintains GRACE legacy.\\ GRACE-FO is a NASA-GFZ collaboration. Includes\\ satellite location data.\end{tabular} & \begin{tabular}[c]{@{}l@{}}Apr 2002- Nov 2009\\ 30 second cadence\\ NASA's Physical\\ Oceanography Distributed\\ Active Archive\\ Center (PO.DAAC)\\ \\ TUDelft \end{tabular} \\ \midrule
 
\begin{tabular}[c]{@{}l@{}}\textbf{SWARM-A\&B}\\ Ground Truth\end{tabular}                                                      & \begin{tabular}[c]{@{}l@{}}Alpha, Bravo, and Charlie (A, B, and C) are identical\\ satellites launched into near-polar low Earth orbit\\ simultaneously for the swarm project.SWARM-A,-B,\\ and -C precess at different local times. For example, \\ SWARM-A and -B precess through 12 hours of local\\ time in 133 days, whereas SWARM-C precesses through\\ 12 hours in 144 days, resulting in a local time gap of 10\\  hours after 5 years of lunch.\end{tabular}                                                             & \begin{tabular}[c]{@{}l@{}}European Space Agency\\ (ESA) Swarm Mission\\ \\ \\ FTP from TUDelft\end{tabular}                                                                 \\ \midrule
 
\begin{tabular}[c]{@{}l@{}}\textbf{10.7}\\ Solar Proxies\end{tabular}                                                           & \begin{tabular}[c]{@{}l@{}}F, M, S, \& Y proxies\\ F10.7 - Flux of solar radiation at 10.7cm wavelength.\\ A long running index. M10.7 - Derived from the Mg II\\ core-to-wing ratio, recording middle-UV near 280nm\\ S10.7 - integrated 26-34nm irradiance measured by\\ the Solar Extreme-UV Monitor above Solar and \\ Heliospheric Observatory Y10.7 - The normalised\\ 81-day F10.7 defines a weighted sum of a solar X-ray \\ region without flare and Lyman-alpha.\end{tabular}                                           & \begin{tabular}[c]{@{}l@{}}Space Environment\\ Technologies\end{tabular}                                                                                                     \\ \midrule
 
\begin{tabular}[c]{@{}l@{}}\textbf{Dst \& Ap}\\ Geomagnetic Proxies\end{tabular}                                                & \begin{tabular}[c]{@{}l@{}}Ap index - Enumerated Kp index, a 3 hour\\ quasi-logarithmic index of magnetic activity relative to\\ a calm day curve. Disturbance Storm Time (Dst) index -\\ measure of geomagnetic activity from observatories\\ near the equator, intensity of the globally symmetrical \\ equatorial electrojet.\end{tabular}                                                                                                                                                                                     & Celetrack \\ \bottomrule
\end{tabular}
\end{table}

\section{Models}
\label{sec:app-models}

Here we describe the embedding model, proxy prediction model, and thermospheric density model proposed here.

\paragraph{VAE}

\begin{figure}
    \centering
    \includegraphics[width=0.5\textwidth]{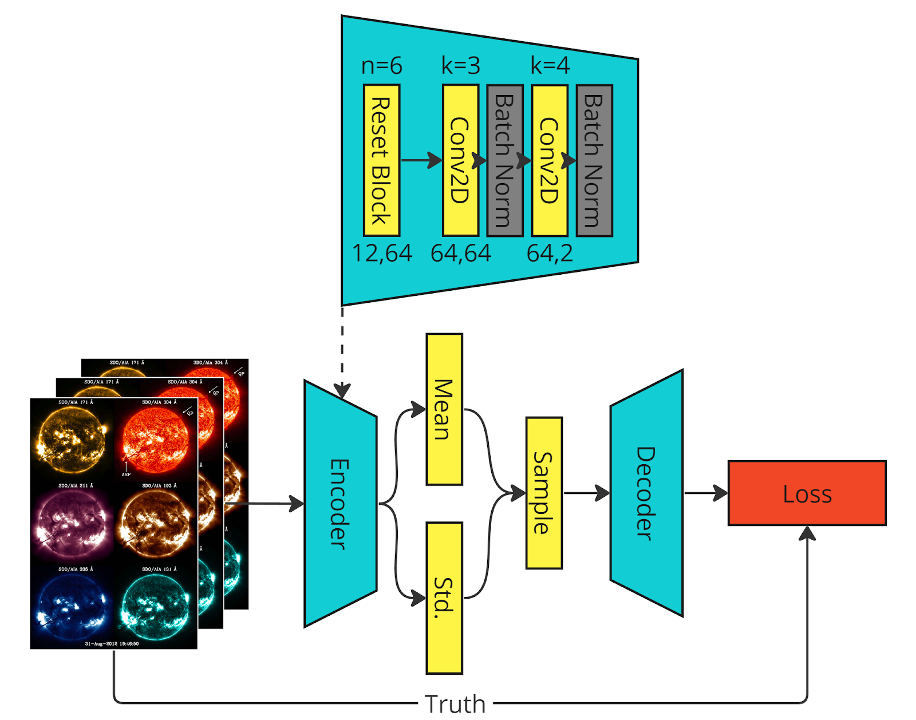}
    \caption{ResNetVAE architecture and training.}
    \label{fig:vae}
\end{figure}

The model architecture and training pipeline is shown in Figure \ref{fig:vae}. We trained this architecture for ten epochs using the Adam optimizer \cite{kingma2014adam} and a learning rate of 0.001. We used a latent embedding size of 4,096 in this work and a validation size of 0.1\%, roughly 2 days.

\paragraph{Proxy indices prediction}

Model hyperparameters are given in Table \ref{tab:f10-hypers}. We train/validation/test split by month as discussed in Section \ref{sec:methods}. 

\paragraph{SDO thermospheric density model}

We experimented with a range of time resolutions and histories. Including more SDOML history and resolution did not significantly improve results, so we opt to include 4 hours of history at a 12 minute resolution such that short term solar activity such as flares are able to be captured by the model \cite{le2015global}. We use mean pooling to aggregate the embeddings. We additionally tried other aggregation methods such as max pooling, a learnable linear combination, and attention-weighted pooling, but these did not significantly improve performance. Model hyperparameters are given in Table \ref{tab:td-hypers}. 

\begin{table}[h]
\caption{Hyperparameter configuration for F10.7 prediction model.}
\begin{tabular}{@{}p{0.35\linewidth}p{0.64\linewidth}@{}}
\toprule
\textbf{Hyperparameter}                        & \textbf{Value}                                                                 \\ \midrule \midrule
\textbf{Optimization}                          &                                                                       \\ \midrule 
Loss Function                            &  Mean Squared Error (MSE)                                                                \\ 
Target                           & F10.7                                      \\ 
Optimizer                             & Adam \cite{kingma2014adam}                                                                               \\
Learning Rate                         & 0.0001                                                                  \\
Batch Size                            & 512                                                                   \\
Early Stopping Patience (epochs)               &   10                                      
                                                                \\
Max Epochs                            & 100                                                                 \\ 
\midrule \midrule
\textbf{Architecture}                          &                                                                       \\ \midrule 
Hidden Layer Non-linearities          & Leaky-ReLU (slope=0.01)                                               \\

Encoder Hidden Dimensions          &   [512,256,256]                                       \\
SDO pooling           & Mean                                                              \\ 
\midrule \midrule
\textbf{Data}                                  &                                                                       \\ \midrule 
SDO Embeddings         & VAE latent mean (4,096 dimensional)                                                                \\
SDO Time Resolution          & 1 hour                                                                  \\
SDO Sequence History          & 1 day                                                                 \\
\bottomrule
\end{tabular}
\label{tab:f10-hypers}
\end{table}

\begin{table}[h]
\caption{Hyperparameter configuration for SDO thermospheric density model.}
\begin{tabular}{@{}p{0.35\linewidth}p{0.64\linewidth}@{}}
\toprule
\textbf{Hyperparameter}                        & \textbf{Value}                                                                 \\ \midrule \midrule
\textbf{Optimization}                          &                                                                       \\ \midrule 
Loss Function                            &  Mean Squared Error (MSE)                                                                \\ 
Target                           &  [Log density] or [log(density) - log(baseline model prediction)]                                                             \\ 
Optimizer                             & Adam \cite{kingma2014adam}                                                                               \\
Learning Rate                         & 0.0001                                                                  \\
Batch Size                            & 64                                                                   \\
Early Stopping Patience (epochs)               & 2                                       
                                                                \\
Max Epochs                            & 4                                                                  \\ 
\midrule \midrule
\textbf{Architecture}                          &                                                                       \\ \midrule 
Hidden Layer Non-linearities          & Leaky-ReLU (slope=0.01)                                               \\
Empirical Features Encoder Hidden Dimensions          &   [256,256,256]                                       \\
Combined Encoder Hidden Dimensions          &   [256,256,256]                                       \\
SDO pooling           & Mean                                                              \\ 
\midrule \midrule
\textbf{Data}                                  &                                                                       \\ \midrule 
SDO Embeddings         & VAE latent mean (4,096 dimensional)                                                                \\
SDO Time Resolution          & 12 minutes                                                                  \\
SDO Sequence History          & 4 hours                                                                  \\
\bottomrule
\end{tabular}
\label{tab:td-hypers}
\end{table}

\section{Effects of short-term EUV irradiance are difficult to observe in the ground truth density data}
\label{sec:app-short-term}

\begin{figure}[h]
    \centering
    \includegraphics[width=0.6\textwidth]{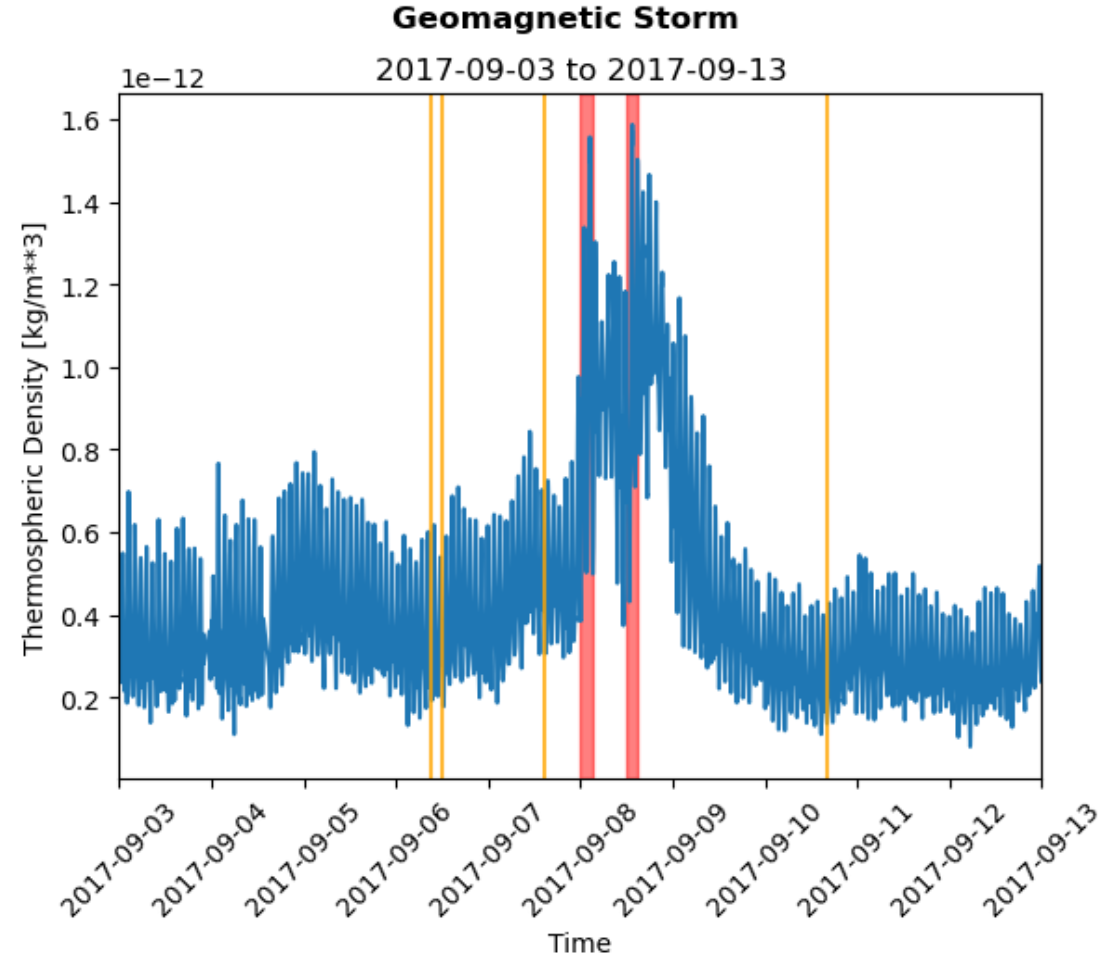}
    \caption{Time series ground truth density during flares and geomagnetic storm periods. Solar flares (orange), periods where the Ap geomagnetic index was 8 or greater (red).}
    \label{fig:geo-storm}
\end{figure}

Figure \ref{fig:geo-storm} shows ground truth densities during a flare event and geomagnetic storm. We find that in many cases we cannot see the flare in the underlying noise, and hence it is difficult to ascertain whether our model captures this short-term activity. Large (X-flare) events without the presence of CMEs or a geomagnetic storm are rare, and these are the only type of short-term variation that will significantly change thermospheric density and thus be relevant for satellite operations \cite{briand2021solar}. 

Nevertheless, by including high-cadence SDO images, our model is theoretically capable of capturing this. Up-sampling flare events may be necessary for the model to adequately learn from these outliers though. An additional by-product of this observation is that changing the history length and time resolution of SDO images used for prediction did not change accuracies significantly.

\end{document}